\documentclass[prd,superscriptaddress,twocolumn,aps,showpacs,amsmath,amssymb,floatfix]{revtex4-2}

\usepackage{color}
\usepackage{bm}
\usepackage{hyperref}
\usepackage{graphicx}
\usepackage{braket}
\usepackage{MnSymbol}
\usepackage{amssymb}
\usepackage{gensymb}
\hypersetup{colorlinks=true}
\usepackage{dcolumn}
\usepackage{amsmath}
\usepackage{amsfonts}
\usepackage{bm}
\usepackage{color}

\begin{document}

\title{Evidence for Isotropic s-Wave Superconductivity in High-Entropy Alloys}

\author{Casey~K.W.~Leung}
\affiliation{Department of Physics, The Hong Kong University of Science and Technology, Clear Water Bay, Kowloon, Hong Kong SAR}
\author{Xiaofu~Zhang}
\affiliation{State Key Laboratory of Functional Materials for Informatics, Shanghai Institute of Microsystem and Information Technology, Chinese Academy of Sciences (CAS), Shanghai 200050, China}
\affiliation{CAS Center for Excellence in Superconducting Electronics, Shanghai 200050, China}
\author{Fabian~von~Rohr}
\affiliation{Department of Chemistry, Universi\"at Z\"urich, Winterthurerstrasse 190, 8057 Zurich, Switzerland}
\author{Rolf.~W.~Lortz}
\affiliation{Department of Physics, The Hong Kong University of Science and Technology, Clear Water Bay, Kowloon, Hong Kong SAR}
\affiliation{IAS Center for Quantum Technologies, The Hong Kong University of Science and Technology, Clear Water Bay, Kowloon, Hong Kong SAR}
\author{Berthold~J\"ack}
\email[]{Correspondence should be addressed to bjaeck@ust.hk}
\affiliation{Department of Physics, The Hong Kong University of Science and Technology, Clear Water Bay, Kowloon, Hong Kong SAR}
\affiliation{IAS Center for Quantum Technologies, The Hong Kong University of Science and Technology, Clear Water Bay, Kowloon, Hong Kong SAR}

\date{\today}

\begin{abstract}
High-entropy alloys (HEA) form through the random arrangement of five or more chemical elements on a crystalline lattice. Despite the significant amount of resulting compositional disorder, a subset of HEAs enters a superconducting state below critical temperatures, $T_{\rm c}<10\,$K. The superconducting properties of the known HEAs seem to suffice a Bardeen–Cooper–Schrieffer (BCS) description, but little is known about their superconducting order parameter and the microscopic role of disorder. We report on magnetic susceptibility measurements on films of the superconducting HEA (TaNb)$_{1-x}$(ZrHfTi)$_{x}$ for characterizing the lower and upper critical fields $H_{\rm c,1}(T)$ and $H_{\rm c,2}(T)$, respectively as a function of temperature $T$. Our resulting analysis of the Ginzburg-Landau coherence length and penetration depth demonstrates that HEAs of this type are single-band isotropic s-wave superconductors in the dirty limit. Despite a significant difference in the elemental composition between the $x=0.35$ and $x=0.71$ films, we find that the observed $T_{\rm c}$ variations cannot be explained by disorder effects.
\end{abstract}

\maketitle

{\em Introduction.--}High-entropy alloys (HEAs) are a new type of alloy with five or more chemical elements arranged on a pseudocrystalline lattice \cite{s1,s2,s3,s4,s5,s6}. A high mixing-entropy minimizes the Gibbs free-energy and facilitates their crystallization on simple lattice structures, such as body-centered cubic (bcc) structure \cite{s3,s4,s5,s6}. Despite the significant amount of compositional disorder, a subset of the HEAs enters a type-II superconducting phase at cryogenic temperatures \cite{s15, s13}. Their large critical fields combined with superior mechanical and thermal properties \cite{s9,s10,s11,s12,s13} render HEA promising candidates for materials applications under extreme conditions, for the fabrication of superconducting magnets, and for superconducting devices based on HEA thin films \cite{s25}.

Ongoing materials synthesis efforts have extended the family of known superconducting HEAs and developed a phenomenological understanding of their properties \cite{s13, s16,s17,s18,s19,s20,s21,s22}. Various analyses of the upper critical field $H_{\rm c2}$ and heat capacity measurements as a function of temperature $T$ support a conventional Bardeen-Cooper-Schrieffer (BCS) pairing mechanism \cite{s13, s15, vonrohr2016effect, s25} with intermediate-strong coupling \cite{vonrohr2016effect}. A dependence of the superconducting transition temperature $T_{\rm c}$ on the chemical composition, as measured in the number of available valence electrons, and mixing entropy has been established \cite{s13, vonrohr2016effect,s25}. Nevertheless, conclusive experimental insight on the superconducting order parameter and the influence of disorder on the superconducting state is missing to date.

\begin{figure}
\includegraphics[width=0.95\linewidth]{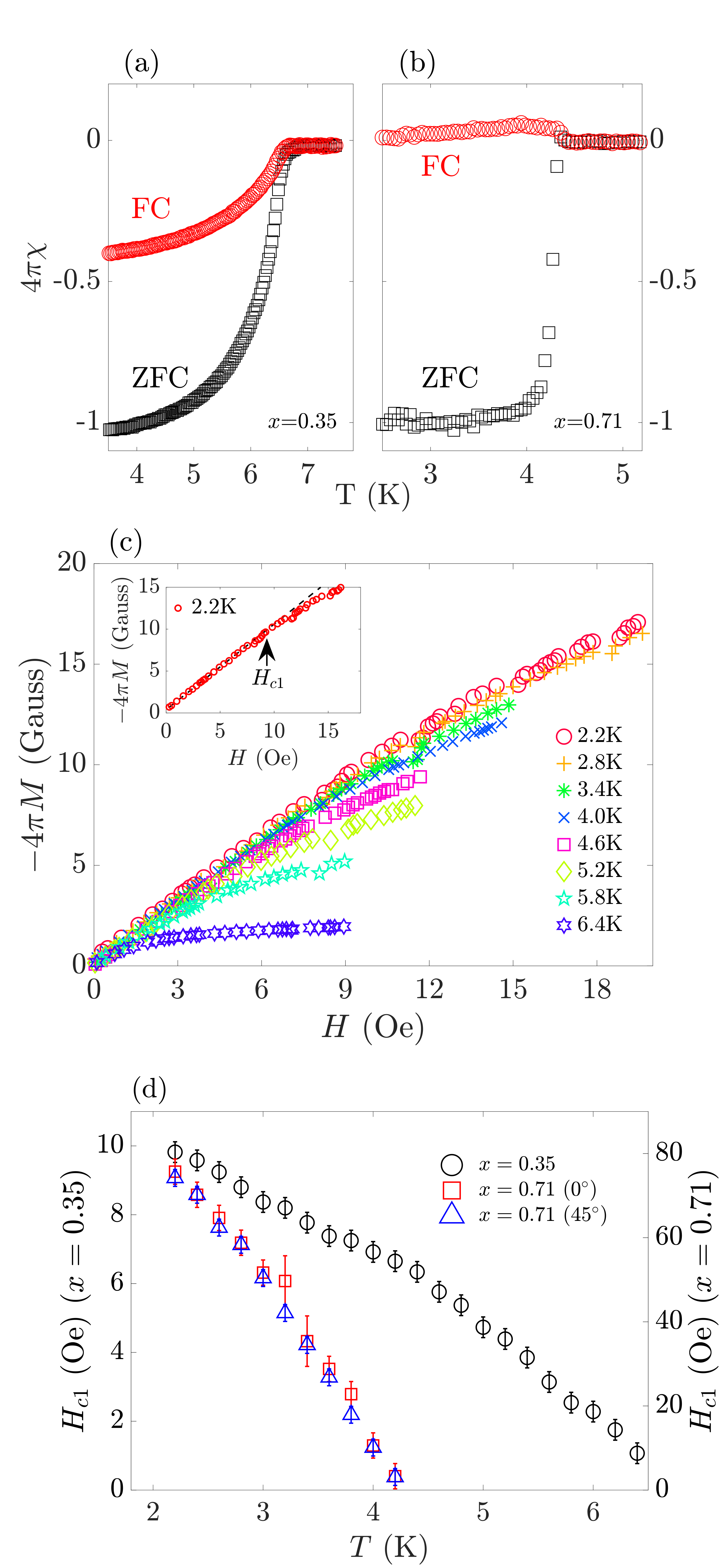}
\caption{ Zero-field cooling (ZFC) and field-cooling (FC) curves of the magnetic susceptibility $\chi$ as a function of temperature, $T$, for the HEA films with $x=0.35$, (a), and $x=0.71$ (b). ZFC (FC) measurements were conducted with $H=5\,$Oe ($H=50\,$Oe) applied in parallel to the film plane. (c) Superconducting volume fraction $-4\pi M$ measured as a function of an external magnetic field $H$ applied in parallel to the $x=0.35$ film at different temperatures indicated in the legend. The inset shows a linear fit to $-4\pi M(H)$ at $T=2.2\,$K for determining the lower critical field $H_{\rm c1}$. The corresponding experimental data for the $x=0.71$ film and details of the procedure for determining $H_{\rm c1}$ are included in Ref.~\cite{SI}. (d). Fitted $H_{\rm c1}(T)$ of the $x=0.35$ (left axis) and $x=0.71$ (right axis) films. Measurements with the magnetic field applied along two different in-plane angles, $0\degree$ and $45\degree$, are shown for the $x=0.71$ data.}
\label{fig:figure1}
\end{figure}

Intuitively, the presence of atomic-scale disorder potentials promotes charge carrier localization in the normal state \cite{s33}. It has long been recognized that a critical amount of disorder can suppress superconductivity near the Anderson quantum phase transition \cite{Ma1985}. The observed decrease in $T_{\rm c}$ with an increase in the mixing entropy, $\Delta S$ seems in agreement with that picture \cite{vonrohr2016effect}. On the other hand, recent analyses of competing interaction-channels in strongly disordered systems predict an enhancement of $T_{\rm c}$ within the BCS framework, when the electron system is tuned to a quantum critical point \cite{Burmistrov2012, Burmistrov2021}. 

This hypothesis finds support in a recent study of HEA $(\rm{TaNb})_{\rm 1-x}(\rm{HfZrTi})_{\rm x}$ thin films; while each of the binary alloys $\rm{TaNb}$ and $\rm{HfZrTi}$ does not show a superconducting transition at or above 2\,K, all solid solutions of $(\rm{TaNb})_{\rm 1-x}(\rm{HfZrTi})_{\rm x}$ at various mixing ratios $x$ are superconducting at $T_{\rm c}\leq6.9$\,K \cite{s25}. The observation of this 'cocktail' effect suggests an intricate relation between the presence of strong compositional disorder, realized through the random arrangement of five atomic species on a bcc lattice, and superconductivity. Therefore, HEA superconductors present a promising test bed for exploring the antagonistic interplay between disorder-driven microscopic charge carrier localization and the formation of a phase-coherent superconducting condensate.

In this letter, we report experimental evidence for conventional superconductivity in the $(\rm{TaNb})_{\rm 1-x}(\rm{HfZrTi})_{\rm x}$ HEAs. To this end, we have performed temperature-dependent magnetization measurements at different alloy compositions $x$. Our analysis of the superconducting penetration-depth $\lambda(T)$ is in quantitative agreement with BCS theory for an isotropic single-band s-wave superconductor in the weak (to intermediate) coupling limit. Our experimental results further show that, despite the large amount of atomic scale disorder, the observed $T_{\rm c}$ variations for films of different elemental compositions do not arise through a disorder driven mechanism.
\\
\\
{\em Experiment.--} Films of superconducting (TaNb)$_{1-x}$(ZrHfTi)$_{x}$ with nominal $x=0.40$ and $x=0.75$ have been prepared by magnetron sputtering on the surface of SiN wafers as described in Ref.~\cite{s25}. (TaNb)$_{1-x}$(ZrHfTi)$_{x}$ forms a single-phase HEA and crystallizes on a pseudo bcc lattice. The pseudo bcc structure, the film thickness ($d\approx1\,\rm{\mu m}$), and the correct stoichiometric composition of the films have been characterized using x-ray diffraction, scanning electron microscopy, and energy-dispersive X-ray spectroscopy measurements, respectively \cite{SI}. The actual compositions $x=0.35$ and $x=0.71$ closely match the targeted compositions. 

\begin{figure}
\includegraphics[width=1\linewidth]{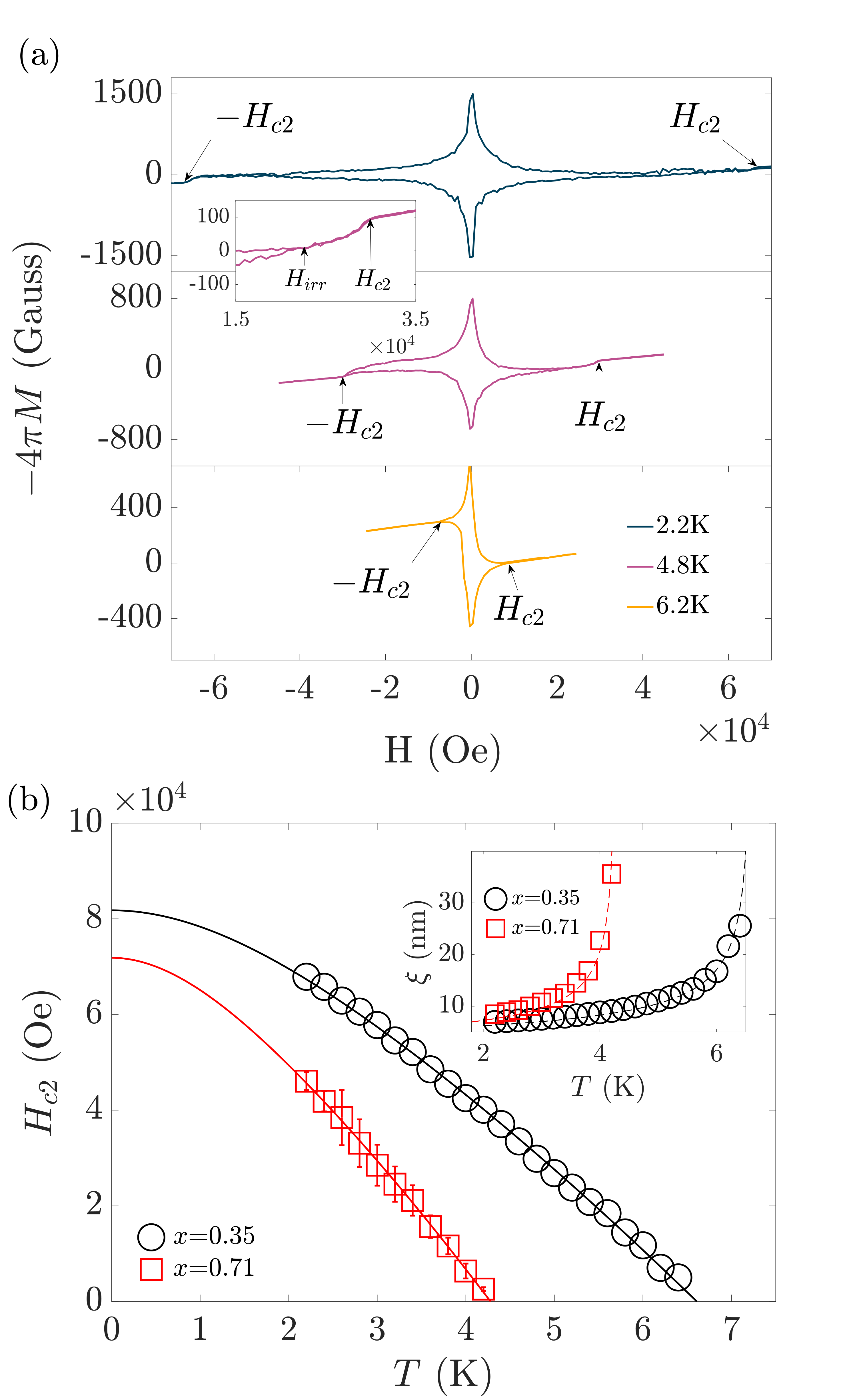}
\caption{(a) Measurements of the superconducting volume fraction $-4\pi M$ as a function of the external magnetic field $H$ applied perpendicular to the sample plane for the $x=0.35$ film. Shown are three representative measurements at indicated temperature, $T$. The complete temperature dependence of the $x=0.35$ film and the corresponding data for the $x=0.71$ film can be found in Ref.~\cite{SI}. (b) Extracted temperature dependence of the upper critical field $H_{\rm c2}(T)$ of the $x=0.35$ and $x=0.71$ films. The solid lines show the fits of $H_{\rm c2}(T)$ with the Werthamer-Helfland-Hohenberg model. \cite{werthamer1966}. The inset shows the corresponding temperature dependence of the Ginzburg-Landau coherence length $\xi$ of the $x=0.35$ and $x=0.71$ films. The dashed lines are the corresponding fits to $\xi(T)$ for superconductors in the dirty limit.}
\label{fig:figure2}
\end{figure}

The molar mixing entropy $\Delta S=R\sum x_{\rm i}\log (x_{\rm i})$ ($R$ - ideal gas constant), of the $x=0.71$ alloy, $\Delta S_{\rm 0.71}=-1.56\,R$, is comparable to that of the $x=0.35$ alloy, $\Delta S_{\rm 0.35}=-1.53\,R$. It is interesting to note that the (TaNb)$_{1-x}$(ZrHfTi)$_{x}$ alloys with similar mixing ratios $x=0.40$ and $x=0.75$ were reported to show the highest $T_{\rm c}\leq7$\,K and highest $H_{\rm c}\approx10\,$T, respectively \cite{s25}. We have performed vibrating sample magnetometry with a commercial superconducting quantum interference device from Quantum Design under cryogenic conditions, $T\geq1.8\,$K for characterizing the superconducting state of the HEA samples. Measuring their magnetic susceptibility $\chi(T)$ we have determined their $T_{\rm c}$, the lower $H_{\rm c1}$ and upper $H_{\rm c2}$ critical fields as a function of temperature and externally applied magnetic field $H$.
\\
\\
{\em Results.--} Zero-field cooling/Field cooling (ZFC/FC) measurement were performed to establish superconductivity in the HEA films, see Fig.~\ref{fig:figure1}(a) and (b). Both the $x=0.35$ and $x=0.71$ film show a diamagnetic response with unity superconducting volume fraction in ZFC measurements at $T\ll T_{\rm C}$. The extracted $T_{\rm c}=(6.7\pm0.1)\,$K and $T_{\rm c}=(4.3\pm0.1)\,$K of the $x=0.35$ and $x=0.71$ film, respectively are in agreement with previous reports \cite{s26}. The FC measurements further indicate strong flux pinning. The diamagnetic response of the $x=0.35$ film is suppressed by about 60\,$\%$, whereas the $x=0.71$ film exhibits a small paramagnetic Meissner effect \cite{braunisch1992}.

The penetration depth of a superconductor can be determined through measurements of $H_{\rm c1}$ and $H_{\rm c2}$. We have determined $H_{\rm c1}(T)$ by mapping out the field response of the HEA films at small external magnetic fields applied in parallel to the film plane. In Fig.~\ref{fig:figure1}(c), we plot the corresponding superconducting volume fraction $-4\pi M(H)$ at different experimental temperatures for the $x=0.35$ film (see Ref.~\cite{SI} for corresponding data of the $x=0.71$ film). At small applied fields, $-4\pi M(H)$ exhibits a linear dependence with a slope of unity. This observation is consistent with the Meissner effect from bulk superconductivity in the HEA films. 

The deviation from linearity at larger $H$ occurs at $H_{\rm c1}$ at which the HEA films enter the mixed phase, i.e., magnetic vortices are penetrating the superconducting volume. We have determined $H_{\rm c1}$ as the field at which the measured $-4\pi M(H)$ data deviate from a linear fit to the small-field region (see Ref.~\cite{SI} for the fitting procedure), see Fig.~\ref{fig:figure1}(c) inset. The resulting $H_{\rm c1}(T)$ is displayed in Fig.~\ref{fig:figure1}(d). While $H_{\rm c1}$ is strongly suppressed for both alloy compositions at $T\rightarrow T_{\rm c}$, $H_{\rm c1}$ of the $x=0.71$ film is about an order of magnitude larger compared to the $x=0.35$ film at $T\ll T_{\rm c}$. Furthermore, other measurements show that $H_{\rm c1}(T)$ is not affected by a 45$\,\%$ rotation of H in the sample plane (see Fig.\,\ref{fig:figure1}(d)).

We have measured the magnetic susceptibility over a larger field range of $-70\,$kOe $<H<+70\,$kOe to further determine the temperature dependence of $H_{\rm c2}$. In Fig.\,\ref{fig:figure2}(a), we plot the corresponding $-4\pi M(H)$ for representative measurements of the $x=0.35$ film (see Ref.~\cite{SI} for the corresponding data of the $x=0.71$ film). We observe a significant magnetic hysteresis between forward and backward sweep, indicative of vortex pinning below the irreversibility field $H_{\rm irr}$ (see inset of Fig.~\ref{fig:figure2}(a)). $H_{\rm c2}$ can be determined from these measurements as the field, at which forward and backward trace deviate from the linear background signal, see marker in Fig.~\ref{fig:figure2}(a). The resulting $H_{\rm c2}(T)$ dependence is shown in Fig.\,\ref{fig:figure2}(b). We observe a monotonic, almost linear, decay of $H_{\rm c2}(T)$ near $T_{\rm c}$ for both alloys.

{\em Discussion.--} We can accurately describe $H_{\rm c2}(T)$ by using the Werthamer-Helfland-Hohenberg (WHH) model of conventional superconductors in the presence of spin-paramagnetism and spin-orbit interaction (see Fig.~\ref{fig:figure2}(b)) \cite{werthamer1966}. Fitting $H_{\rm c2}(T)$, we obtain $H_{\rm c2, 0}=(81.8\pm0.4)\,$kOe and $H_{\rm c2, 0}=(71.9\pm0.6)\,$kOe for the $x=0.35$ and $x=0.71$ film, respectively. These values are significantly smaller than the values of the corresponding Pauli paramagnetic limit in the weak coupling limit $H_{\rm P}=18.4T_{\rm C}$ ($H_{\rm P}$ in kOe and $T_{\rm C}$ in K) \cite{clogston1962, chandrasekhar1962note}. $H_{\rm P}=(123.3\pm1.8)\,$kOe for the $x=0.35$ film and $H_{\rm P}=(79.1\pm1.8)\,$kOe for the $x=0.71$ film, indicating that superconductivity is rather limited by orbital effects induced by the externally applied field. 

We obtain the Ginzburg-Landau (GL) coherence length $\xi$ through the analysis of $H_{c2}=\phi_{0}/(2\pi{\xi}^2)$. ${\phi}_0=h/2e=2.07\times 10^{-7}\,$Oe cm$^2$ corresponds to the magnetic flux quantum, $h$ to Planck's constant, and $e$ to the electron charge. The resulting $\xi(T)$ are shown in the inset of Fig.\,\ref{fig:figure2}(b) for both alloy compositions. Their diverging characteristics for $T\rightarrow T_{\rm c}$ satisfies the GL description of conventional superconductors in the dirty limit $\xi=0.855\sqrt{\xi_{\rm 0}l}/\sqrt{1-T/T_{\rm C}}$. $l$ denotes the electron mean free path and in the dirty limit $\xi\approx l$. $\xi_{\rm 0}=\sqrt{\phi_{0}/2\pi H_{\rm c2, 0}}$ can be calculated from the WHH analysis, $\xi_{\rm 0,\,x=0.35}=(6.30\pm0.01)\,$nm and $\xi_{\rm 0,\,x=0.71}=(6.80\pm0.01)\,$nm. Fitting $\xi(T)$, as shown in the inset of Fig.\,\ref{fig:figure2}(b), we obtain $l_{\rm 0.35}=(5.80\pm0.18)\,$nm and $l_{\rm 0.71}=(5.65\pm0.32)\,$nm for the $x=0.35$ and $x=0.71$ film, respectively. The observation $l_{\rm 0.35}\approx l_{\rm 0.71}$ is consistent with the comparable mixing entropy of both films, i.e., a comparable amount of atomic scale disorder in the samples.

\begin{figure}[t!]
\includegraphics[width=1\linewidth]{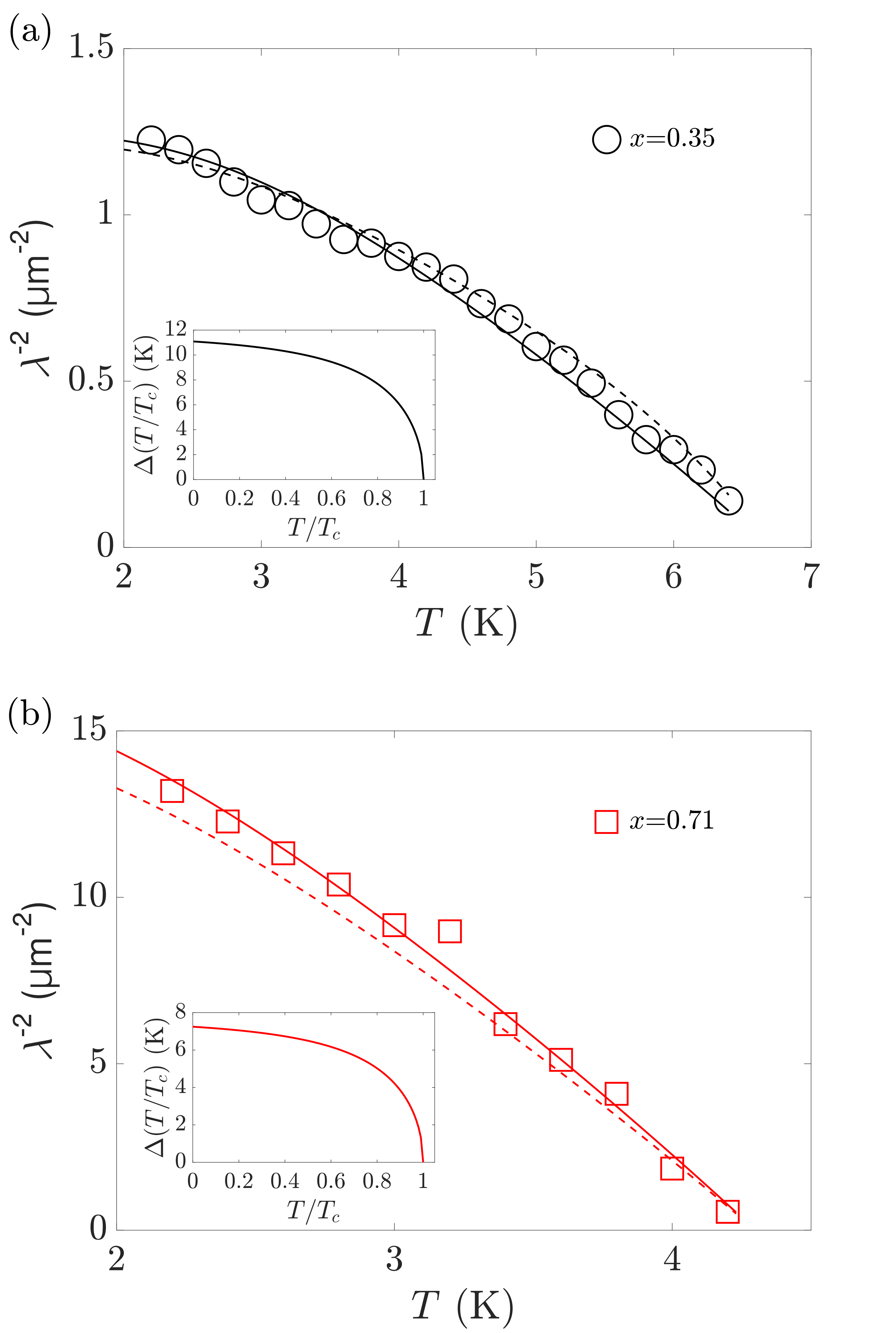}
\caption{Experimentally determined values of the superconducting penetration depth, $\lambda(T)$, of the (a) $x=0.35$ and (b) $x=0.71$ films are plotted as $\lambda^{-2}$ as a function of temperature, $T$ (open symbols). The solid (dashed) lines are the corresponding fits to the data using Eq.~\ref{EQ1} with $\alpha=1.74$ ($\alpha=2.2$). (c) The temperature-dependent quasiparticle gaps $\Delta(T)$ of the $x=0.35$ and $x=0.71$ films obtained from fitting the data in (a) and (b) are shown in the respective insets}
\label{fig:figure3}
\end{figure}

Experimental values of $\xi(T)$ and $H_{c1}(T)$ can be used to determine $\lambda(T)$ by using the relation ${\mu}_0 H_{c1}={\phi}_0/{4\pi\lambda^2}\ln(\lambda/\xi)$. Knowledge of $\lambda(T)$ can provide valuable insights into the nature of superconductivity in the HEA films. We have analyzed $\lambda^{-2}(T)$ of both films, which are shown in Fig.\,\ref{fig:figure3}(a) and (b), by using the superfluid density model of conventional superconductors in the framework of the BCS theory \cite{tinkham2004introduction}, 
\begin{equation}
    \rho(T)=\frac{\lambda(0)^{2}}{\lambda(T)^{2}}=1+2\int^\infty_{\Delta(T)} dE \frac{\partial f}{\partial E} \frac{E}{\sqrt{E^2-\Delta(T)^2}}.
    \label{EQ1}
\end{equation}
$f=[1+\exp(E/k_{\rm B} T)]^{-1}$ denotes the Fermi function ($E$ - energy; $k_{\rm B}$ - Boltzman's constant) and $\Delta(T)$ the temperature-dependent superconducting quasiparticle gap). 

The isotropic response of the superconducting state to an external magnetic field, cf. Fig.\,\ref{fig:figure1}(d), supports an isotropic superconducting order parameter. Therefore, we assume superconductivity with an isotropic s-wave symmetry and single-band pairing for our analysis. The corresponding interpolating BCS gap function reads $\Delta(T)=\Delta_0 \tanh({\alpha\sqrt{T_{\rm C}/T-1}})$ with $\Delta_{\rm 0}=1.764k_{\rm B}T_{\rm C}$. Using this model, we can accurately fit the temperature-dependence of $\lambda^{-2}$ at both alloy compositions, see Fig.\,\ref{fig:figure3}(a). Fitting results in $\lambda_{\rm x=0.35}(0)=(896\pm4)\,$nm and $\lambda_{\rm x=0.71}(0)=(245\pm2)\,$nm as the penetration depth in the zero temperature limit. The calculated quasiparticle gaps, which were used for fitting $\lambda^{-2}(T)$, are displayed in the insets of Fig.~\ref{fig:figure3}(a) and (b). $\lambda^{-2}(T)$ of both films is in agreement with the weak-coupling BCS limit, $\alpha_{\rm0.35}=\alpha_{\rm0.71}=1.74$. Nevertheless, we also observe good agreement between experiment and model at intermediate coupling $\alpha=2.2$, which is consistent with previous reports from heat capacity measurements \cite{vonrohr2016effect}. Overall, our analysis shows that the (TaNb)$\rm_{1-x}$(ZrHfTi)$\rm_x$ HEAs are single-band isotropic s-wave superconductors that can be described by BCS theory.

The large degree of compositional disorder is expected to result in a significant on-site potential disorder at the atomic scale \cite{jack2021visualizing}. Therefore, the strong dependence of $T_{\rm c}$ on $x$, $T_{\rm c,\,x=0.35}=(6.7\pm1.1)\,$K and $T_{\rm c, \,x=0.71}=(4.3\pm1.1)\,$K, invites speculation on the role of disorder for $T_{\rm c}$ and the superconducting mechanism more generally \cite{Ma1985, Burmistrov2012, Burmistrov2021}. Our analysis reveals a comparable mean free path on the order of 6\,nm at both alloy compositions (see Fig.~\ref{fig:figure2}(b)). This observation is consistent with an almost equivalent mixing entropy, despite their different elemental composition. Even though both films fall into the dirty limit of superconductivity at $\xi\approx l$, our results do not thus support a connection between disorder and $T_{\rm c}$ variations. Instead, the isotropic s-wave superconductivity observed at both alloy compositions points to a different picture, in which the observed $T_{\rm c}$ variations arise within the BCS framework, i.e., from changes to the density of states at the Fermi level induced by electronic doping \cite{vonrohr2016effect}.

{\em Note}: All measured and fitted values of the relevant quantities are tabulated in Ref.~\cite{SI} and the raw data and corresponding analysis are available through Ref.~\cite{repository}.

{\em Conclusion.--} 
We have experimentally studied the superconducting state of the HEA (TaNb)$_{1-x}$(ZrHfTi)$_{x}$ films with $x=0.35$ and $x=0.71$ by measuring  $H_{\rm c1}(T)$ and $H_{\rm c2}(T)$. Our analysis of $\lambda(T)$ is in quantitative agreement with the BCS theory of an isotropic single band s-wave superconductor in the weak (to intermediate) coupling limit. The analysis of $\xi(T)$ reveals a comparable amount of disorder at both compositions, $l_{\rm x=0.35}\approx l_{\rm x=0.71}$. Therefore, we can exclude that the observed variations in $T_{\rm c}$ originate from a disorder-driven mechanism. Further theoretical and experimental studies will be needed for characterizing the low-energy electronic structure at various alloy compositions and its influence on $T_{\rm c}$.

Looking ahead, results of such efforts may inform pathways for realizing HEAs with enhanced superconducting $T_{\rm c}$. Employing penetration-depth measurements to other superconducting HEAs \cite{s13}, such as those crystallizing on the CsCl-type lattice, it will be interesting to test whether weak coupling s-wave superconductivity is a common occurrence in these material systems. While the $T_{\rm c}$ variations of bulk superconductivity appear to be independent from disorder, the study of these or other HEAs films in the two-dimensional limit with maximized on-site disorder could offer avenues for exploring $T_{\rm c}$ enhancements through multifractal eigenstates near a quantum critical point \cite{evers2008anderson, Burmistrov2012, Burmistrov2021}

\bibliography{literature}

\end{document}